\documentclass{appolb}
\pdfoutput=1 
\usepackage{graphicx}
\usepackage{amssymb}
\usepackage{amsmath}
\usepackage[utf8x]{inputenc}
\usepackage[T1]{fontenc} 
\usepackage{physics}
\newcommand{\del}{\partial}
\usepackage{bm}
\usepackage{bbm}
\usepackage{multirow}
\usepackage[table,xcdraw]{xcolor}
\usepackage{mathtools}
\usepackage{nccmath}
\usepackage{dsfont}
\usepackage{float}
\usepackage{caption}
\usepackage{subcaption}
\usepackage[symbol]{footmisc}
\usepackage{xpatch}

\newcommand{\be}{\begin{equation}}
	\newcommand{\ee}{\end{equation}}
\newcommand{\bea}{\begin{array}}
	\newcommand{\ea}{\end{array}}
\newcommand{\beqa}{\begin{eqnarray}}
	\newcommand{\eeqa}{\end{eqnarray}}
\newcommand{\nn}{\nonumber}

\usepackage{perpage}
\MakePerPage{footnote}

\begin{document}
		
	\title{Thermalization in massive deformations of Yang-Mills matrix models
	}
	\author{Onur Oktay\footnote{E-mail: oktay24005@gmail.com}
	}
	\maketitle
	\begin{abstract}
		We numerically study 
		the classical evolution of a Yang-Mills matrix model with two distinct mass deformation terms, which can be contemplated as a massive deformation of the bosonic part of the BFSS model. 
		Through numerical analysis, it is shown that 
		when the simulations are started from a certain set of initial conditions, thermalization occurs. Besides, an estimation method is proposed to determine the approximate thermalization time.
		Using this method, we demonstrate that
		thermalization time vary logarithmically with 
		increasing matrix size when the mass terms differ. Introducing a matrix configuration, we also obtain reduced actions and subsequently analyze how the thermalization time change as a function of the energy.
	\end{abstract}

	\section{Introduction}
	
	Since the introduction of the gauge/gravity duality \cite{Maldacena:1997re, Witten:1998qj}, there has been an immense effort from the theoretical physics community to use holographic methods in order to gain enhanced understanding about various physical phenomena. The duality between a thermal state in the boundary theory and a black hole in the bulk has formed the backbone of 
	these studies
	and enabled the researchers to associate the process of thermalization in a unitary field theory with the formation of a black hole in the dual side. Explaining the dynamics of thermalization in isolated quantum systems, which is a central problem in many-body physics \cite{srednicki:1994, deutsch:1991}, has also been studied within the context of AdS/CFT 
	by focusing on certain string theory-inspired constructions such as the BFSS \cite{Banks:1996vh} and BMN \cite{Berenstein:2002jq} matrix models. 
	
	Over a decade ago, the remaining mysteries about 
	the nature 
	of black holes have led to several speculations related to their quantum mechanical structure, some of which could be tested in matrix model environments. 
	To elaborate, motivated by the arguments of \cite{Hayden:2007cs}, Sekino and Susskind have conjectured that black holes are fast scramblers i.e. they scramble information at a rate proportional to the logarithm of the number of degrees of freedom \cite{Sekino:2008he}.
	The thermalization processes observed in the BMN model have been numerically investigated and reported in a series of papers \cite{Asplund:2011qj, Berenstein:2010bi, Asplund:2012tg}. 
	In particular, 
	the results obtained in \cite{Asplund:2011qj} are intriguing as they are broadly consistent with the fast scrambling conjecture. Berenstein et al. have shown that 
	simulations of thermalization in the BMN model
	provide numerical evidence for fast thermalization, which may also be interpreted to implicate fast scrambling. 
	
	On the other hand, extensive thermodynamic simulations of the BFSS model, 
	including detailed numerical studies of thermalization times,
	have been performed in references \cite{Riggins:2012qt,Aoki:2015uha}. 
	Furthermore, the relation between quantum chaos and thermalization has been recently explored in \cite{Buividovich:2018scl}. Besides these developments, it is essential to note that 
	that due to large number of degrees of freedom interacting through a quartic Yang-Mills potential, it does not appear quite possible that general solutions of the BFSS/BMN models can be determined. Even the smallest Yang-Mills matrix model with two $2 \times 2$ matrices and with $SU(2)$ gauge symmetry has not been completely solved until this date \cite{Berenstein:2016zgj}. Thus, in order to reach meaningful results, instead of considering the whole matrix theory it seems reasonable to
	concentrate on simplified structures with less degrees of freedom. 
	A convenient way of achieving this is to
	place prior constraints on the system at hand by starting the 
	simulations with specified sets of initial conditions. Although
	one would ideally prefer to choose initial conditions with the aim of setting up a configuration, which resembles the phenomenon 
	of scattering gravitons at high energies, currently this not possible 
	in the BFSS case due to the insufficient understanding of graviton states in this matrix model \cite{Berenstein:2010bi}. Nevertheless, as it will be discussed shortly, valuable information regarding the thermalization phenomenon can still be gathered from certain gauge invariant massive deformations of the BFSS model.   
	
	In this paper, our main interest is to analyze the dynamics of thermalization in a Yang-Mills matrix model with two distinct mass deformation terms, whose emerging chaotic motions have been investigated in \cite{Baskan:2019qsb}. This model has the same matrix content as the bosonic part of the BFSS matrix model, but also contains mass deformation terms that keep the gauge invariance intact. The paper is structured as follows. Section \ref{secYM} starts out with a brief introduction of the model, which is followed by the description of the initial conditions that are used in the simulations. In section \ref{secNum}, we investigate the thermalization processes observed in the matrix model with massive deformations by performing a detailed numerical analysis of its classical evolution. 
	This is followed by an examination of the variation of thermalization time with respect to matrix size. 
	Then, by introducing a configuration of matrices, 
	we obtain reduced actions from the full matrix model and subsequently  
	explore the change of thermalization time with the energies of these reduced actions. Lastly, section \ref{Concs} is devoted to conclusions and outlook.  
	
	\section{Yang-Mills matrix model with double mass deformation} \label{secYM}
	
	The BFSS matrix model is a Yang-Mills theory in $0+1$ dimensions which arises from the dimensional reduction of the Yang-Mills theory in $9+1$ dimensions with $\mathcal{N}=1$ supersymmetry \cite{Banks:1996vh}. In this paper, we focus upon a
	gauge invariant double mass deformation of the bosonic part of the BFSS action which may be specified as \cite{Baskan:2019qsb}
	\be
	\label{MD}
	S = \frac{1}{g^2} \int dt \, \tr( \frac{1}{2}(D_t B_I)^2 + \frac{1}{4} 
	{\lbrack B_I, B_J \rbrack}^2 - \frac{1}{2} \mu_1^2 B_i^2 - \frac{1}{2} \mu_2^2 B_k^2 ) \,, 
	\ee
	where the indices $i$ and $k$ take on the values $i = 1,2,3$ and $k = 4,5,6$, respectively.  
	In (\ref{MD}), $B_I$ $(I=1,\dots,9)$ are $N \times N$ Hermitian matrices and $ \tr$ stands for the trace. The covariant derivatives are defined by
	\be
	D_t {B}_I=\del_t B_I - i \lbrack A, B_I \rbrack \, .
	\ee
	When the deformation parameters $\mu_1$ and $\mu_2$ are both equal to zero, (\ref{MD}) reduces to the bosonic part of the classical BFSS action. Since, we are going to be essentially concerned with the classical dynamics of~\eqref{MD}, we 
	absorb the coupling constant in the definition of $\hbar$, 
	as it only determines the overall scale of energy classically. 
	
	In the Weyl gauge, $A=0$, the equations of motion for $B_I$ take the form
	\begin{subequations}
		\begin{align}
			\ddot{B_i} + \lbrack B_I , \lbrack B_I , B_i \rbrack \rbrack  + \mu_1^2 B_i &= 0 \,, \\ 
			\ddot{B_k} + \lbrack B_I , \lbrack B_I , B_k \rbrack \rbrack +  \mu_2^2 B_k &= 0 \,, \\
			\ddot{B_r} + \lbrack B_I , \lbrack B_I , B_r \rbrack \rbrack &= 0 \,,
		\end{align}
	\end{subequations} 
	where the index $r$ runs through the values $7,8,$ and $9$.
	Similarly, 
	the Weyl gauge Hamiltonian reads
	\be
	\label{Ham1}
	H = \tr( \dfrac{{P_I}^2}{2} -  \frac{1}{4} 
	{\lbrack B_I, B_J \rbrack}^2 + \frac{1}{2} \mu_1^2 B_i^2 + \frac{1}{2} \mu_2^2 B_k^2 ) \,.
	\ee
	Due to gauge invariance, $B_I$ matrices and conjugate momenta should also satisfy the Gauss Law constraint given by
	\be
	\label{GaussLaw}
	\lbrack B_I, P_I \rbrack = 0 \,.
	\ee
	The Hamilton's equations of motion can easily be derived from (\ref{Ham1}). However, in order to obtain relations that are more convenient for numerical simulations, we rename a subset of phase space coordinates (namely $B_I$ and $P_I$ for $I \geqslant 4$) and subsequently change the indices labelling the aforementioned coordinates so that all indices can range over the same set of integer values. The resulting equations of motion can be written out as follows
	\begin{subequations}
		\label{HeomNw} 
		\begin{align}
			\dot{P_i} &= \lbrack \lbrack B_j , B_i \rbrack, B_j \rbrack + 
			\lbrack \lbrack C_l , B_i \rbrack, C_l \rbrack + 
			\lbrack \lbrack D_s , B_i \rbrack, D_s \rbrack - \mu_1^2 B_i  \,, \label{YMDMeomA1} \\ 
			\dot{R_l} &= \lbrack \lbrack B_i , C_l \rbrack, B_i \rbrack + 
			\lbrack \lbrack C_{l^\prime} , C_l \rbrack, C_{l^\prime} \rbrack + 
			\lbrack \lbrack D_s , C_l \rbrack, D_s \rbrack - \mu_2^2 C_l  \,, \label{YMDMeomB1} \\ 
			\dot{W_s} &= \lbrack \lbrack B_i , D_s \rbrack, B_i \rbrack + 
			\lbrack \lbrack C_l , D_s \rbrack, C_l \rbrack + 
			\lbrack \lbrack D_{s^\prime} , D_s \rbrack, D_{s^\prime} \rbrack \,, \label{YMDMeomC1} \\
			P_i &= \dot{B_i} \,, \quad R_l = \dot{C_l} \,, \quad W_s = \dot{D_s} \,,  
		\end{align}
	\end{subequations}
	where $j,l,l^\prime,s,s^\prime = 1,2,3$. 
	Furthermore, in this new notation (\ref{GaussLaw}) becomes  
	\be
	\label{GaussLawNew}
	G = \lbrack B_i, P_i \rbrack + \lbrack C_l, R_l \rbrack + \lbrack D_s, W_s \rbrack = 0 \,.
	\ee
	
	One of the primary purposes of this study is to examine the dependence of thermalization on the choice of initial conditions. To this end, we adopt an approach similar to the one suggested in \cite{Asplund:2011qj} and set up the initial conditions as follows
	\begin{align}
		\label{IniConds} 
		B_1 &= 
		\begin{pmatrix}
			J_1 & 0  \\
			0 & 0  
		\end{pmatrix} \,, \quad
		B_2 = 
		\begin{pmatrix}
			J_2 & q_1  \\
			{q_1}^\dagger & 0  
		\end{pmatrix} \,, \quad
		B_3 = 
		\begin{pmatrix}
			J_3 & q_2  \\
			{q_2}^\dagger & 0  
		\end{pmatrix} \,,  \quad 
		C_l = 
		\begin{pmatrix}
			J_l & 0  \\
			0 & 0  
		\end{pmatrix}  \nn \\
		P_1 &= 
		\begin{pmatrix}
			0 & 0  \\
			0 & p_0  
		\end{pmatrix} \,, \quad P_2 = P_3 = 0 \,, \quad D_s = 0 \,, \quad R_l = W_s = 0 \,, 
	\end{align}
	where $J_i$'s are  $(N-1)$-dimensional Hermitian matrices. They denote the spin-$j$ 
	$(j=(N-2)/2)$
	irreducible representation of $SU(2)$ and form the fuzzy two-sphere at level $j$ 
	\cite{Madore:1991bw,Balachandran:2005ew}. 
	While the diagonal modes of $B_i$ and $C_l$ matrices start from the fuzzy sphere configurations, $P_1$ initiates with a single eigenvalue, $p_0$, on the main diagonal. Besides their diagonal modes, the off-diagonal elements of $B_2$ and $B_3$ are also excited with the addition of $q_1$ and $q_2$ blocks that are consisting of randomly generated initial conditions. These blocks, which serve as sources of small fluctuations, are formed by utilizing a complex normal distribution with a spread proportional to
	${(\hbar/(N-1))}^\frac{1}{2}$.
	After completing the essential procedure of specifying initial conditions, we may now proceed to the stage of numerical simulations. 
	
	\section{Numerical results} \label{secNum}   
	
	This section is devoted to an investigation of the thermalization processes observed in the Yang-Mills theory with massive deformations. In order to provide a comprehensive analysis, we carry out numerical simulations for the time evolution of (\ref{HeomNw}). 
	After discretizing the equations of motion, an iterative algorithm can be developed to solve the discretized equations numerically.
	By saving the contents of the eighteen matrices every few iterations,
	one can gain valuable insight into the dynamics of thermalization as will be discussed shortly.
	
	In the computations, a simulation code implemented in Matlab is used. The code is executed with a constant time step of $0.004$ and we run it for a sufficient amount of time to clearly observe the values that the eigenvalues converge to. Due to truncation of digits, errors are inevitable in numerical calculations. In this regard, although the initial conditions given by (\ref{IniConds}) fulfill the Gauss law constraint, the cumulative effect of rounding errors could cause the violation of (\ref{GaussLawNew}). However, by constantly monitoring $G$ during the the trial runs of the simulation, 
	we made sure that no such effect is present.         
	
\begin{figure}[!htb]
		\centering
		\begin{subfigure}[!htb]{.32\textwidth}
			\centering
			\includegraphics[width= 1\linewidth]{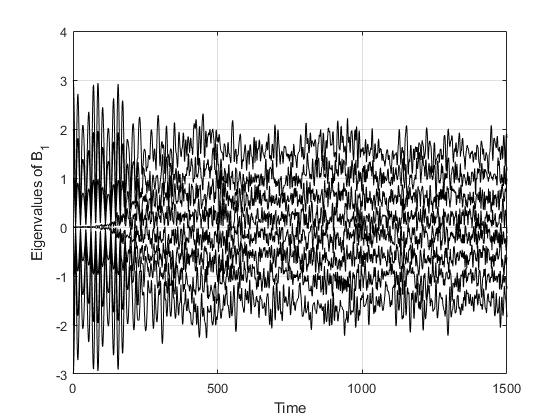}  
			\caption{$p_0 = 0$}
			\label{fig:fig1a1}
		\end{subfigure}	
		\begin{subfigure}[!htb]{.32\textwidth}
			\centering
			\includegraphics[width=1\linewidth]{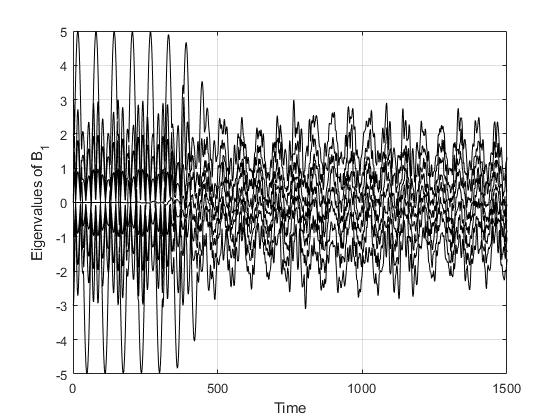}  
			\caption{$p_0 = 5$}
			\label{fig:fig1b1}
		\end{subfigure}	
		\begin{subfigure}{.32\textwidth}
			\centering
			\includegraphics[width=1\linewidth]{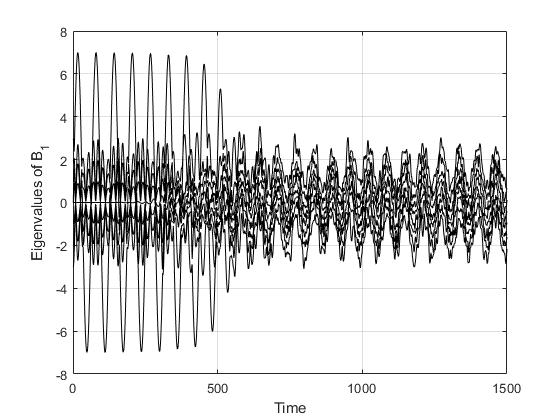}  
			\caption{$p_0 = 7$}
			\label{fig:fig1c1}
		\end{subfigure}	
		\begin{subfigure}{.32\textwidth}
			\centering
			\includegraphics[width=1\linewidth]{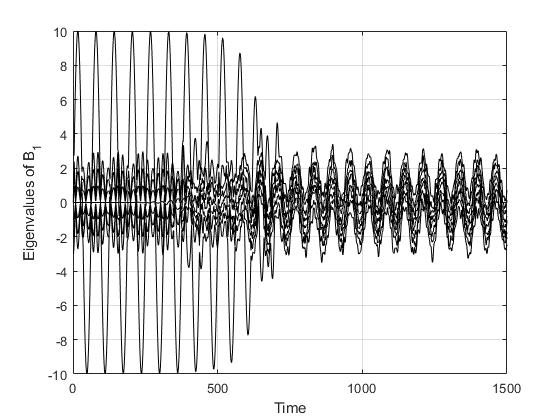}  
			\caption{$p_0 = 10$}
			\label{fig:fig1d1}
		\end{subfigure}	
		\begin{subfigure}{.32\textwidth}
			\centering
			\includegraphics[width= 1\linewidth]{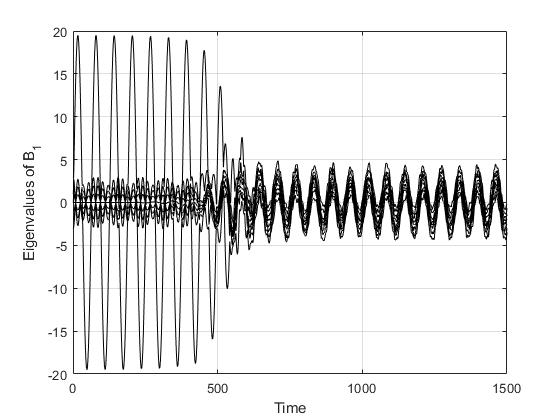}  
			\caption{$p_0 = 19.5$}
			\label{fig:fig1e1}
		\end{subfigure}	
		\begin{subfigure}{.32\textwidth}
			\centering
			\includegraphics[width=1\linewidth]{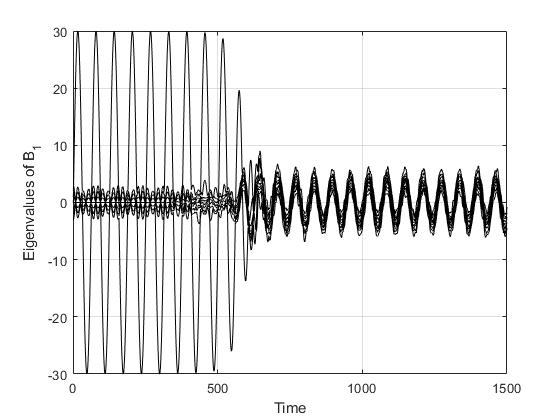}  
			\caption{$p_0 = 30$}
			\label{fig:fig1f1}
		\end{subfigure}	
		\caption{Eigenvalues of $B_1$ vs. Time at $N=8$, $\mu_1=1$, and $\mu_2=1.5$}
		\label{fig:fig11}
	\end{figure}

	Having now introduced the basic features of numerical computations, we move on to the details of obtained results. When the random fluctuation terms are not added to the system, i.e. $\hbar=0$,
	the starting configurations keep evolving periodically in time 
	and thermalization does not occur. Thus, to avoid such a scenario, we set the value of $\hbar$ to $0.001$, which will remain fixed for the rest of this work. In order to discover various intriguing properties of the thermalization process, we first vary the $p_0$ parameter. Figure \ref{fig:fig11} shows the evolution of the eigenvalues of $B_1$ with simulation time for six different $p_0$ values.
	The first thing we can immediately observe from the plots is that the oscillatory behavior of eigenvalues,
	which can be most clearly seen from the last two figures,
	 become more apparent with increasing $p_0$. In Figures \ref{fig:fig1e1} and \ref{fig:fig1f1}, after a series of oscillations, 
	the amplitudes of the oscillations
	decrease considerably and the frequencies tend to synchronize which results in the emergence of collective oscillations.      
	\begin{figure}[!htb]
		\centering
		\begin{subfigure}[!htb]{.48\textwidth}
			\centering
			\includegraphics[width= 1\linewidth]{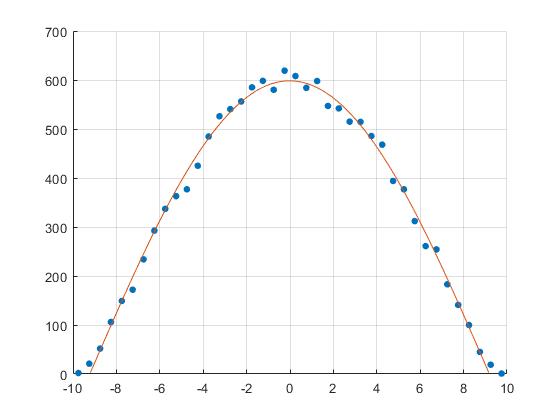}  
			\caption{$p_0 = 19.5$}
			\label{fig:fig2_a}
		\end{subfigure}	
		\begin{subfigure}[!htb]{.48\textwidth}
			\centering
			\includegraphics[width=1\linewidth]{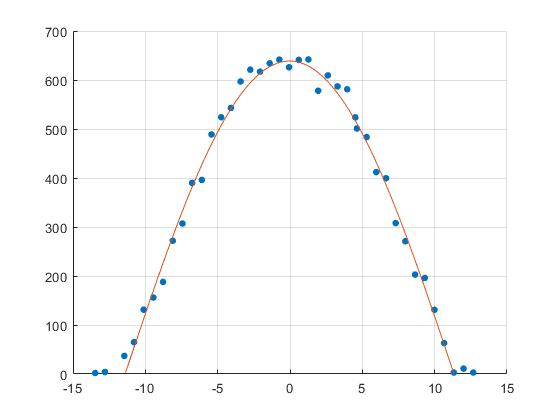}  
			\caption{$p_0 = 30$}
			\label{fig:fig2_b}
		\end{subfigure}	
		\caption{Histograms of eigenvalues of $P_1$ at $N=8$, $\mu_1=1$, and $\mu_2=1.5$}
		\label{fig:fig2_1}
	\end{figure}
	
\begin{figure}[!htb]
		\centering
		\includegraphics[width= 0.7\linewidth]{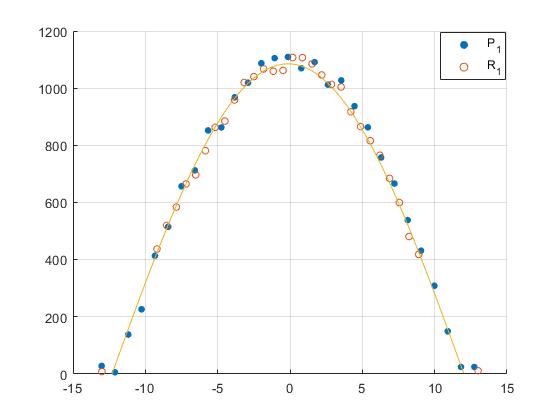}  
		\caption{Histograms of eigenvalues of $P_1$ and $R_1$ at $N=8$, $\mu_1=1$, and $\mu_2=1.5$}
		\label{fig:fig31}
	\end{figure}

	On the other hand, as it is described in detail in subsection \ref{Ssec_Thrm}, thermalization occurs at all six $p_0$ values that are used in preparation of Figure \ref{fig:fig11}. In order to probe the presence of thermalization occuring at the $p_0$ values of $19.5$ and $30$, let us consider the results shown on Figure \ref{fig:fig2_1}. In Figures \ref{fig:fig2_a} and \ref{fig:fig2_b}, the eigenvalue distributions of the momentum matrix $P_1$ are illustrated. The histograms are generated by sampling the eigenvalues of $P_1$ on the time interval\footnote[2]{A detailed discussion of the determination of thermalization times is given in the next subsection.}
	$[758,2500]$ during which 
	the system resides in potentially thermalized states. 
	The bin size is set to 40 and the dots in the figure correspond to the midpoints of the top edges of histogram bars. As expected from thermalized configurations, the semicircle distribution model fits the data nicely in both cases. Furthermore, in order to compare the eigenvalue distributions of the momenta matrices, the histograms of the eigenvalues of $P_1$ and $R_1$ are plotted together in Figure \ref{fig:fig31}. This time the histograms are generated by sampling the eigenvalues on the time interval $[758,3000]$ with a bin size equal to $30$.
	Let us also note that we now set the value of $p_0$ to $30$, which
	will remain fixed for the rest of this work unless otherwise stated.
	It appears   
	that the semicircle model gives an essentially good fit to both $P_1$ and $R_1$ distributions, which implies that after $t = 758$ momenta temperatures become essentially the same. Thus, it is safe to conclude that thermalization has occurred. 
	
	\subsection{Thermalization time} \label{Ssec_Thrm}

	The main results concerning the 
	presence of thermalization  
	have been discussed up to this point. As it is central to the understanding of the thermalization process, let us consider a method that will help us in both determining the thermalization time of the system and providing evidence for the presence of thermalization. This method
	relies on the evaluation of the relative size of changes in both $B_i$ and $C_l$ eigenvalues \cite{Asplund:2011qj}. 
	
\begin{figure}[!htb]
		\centering
		\includegraphics[width= 1.1\linewidth]{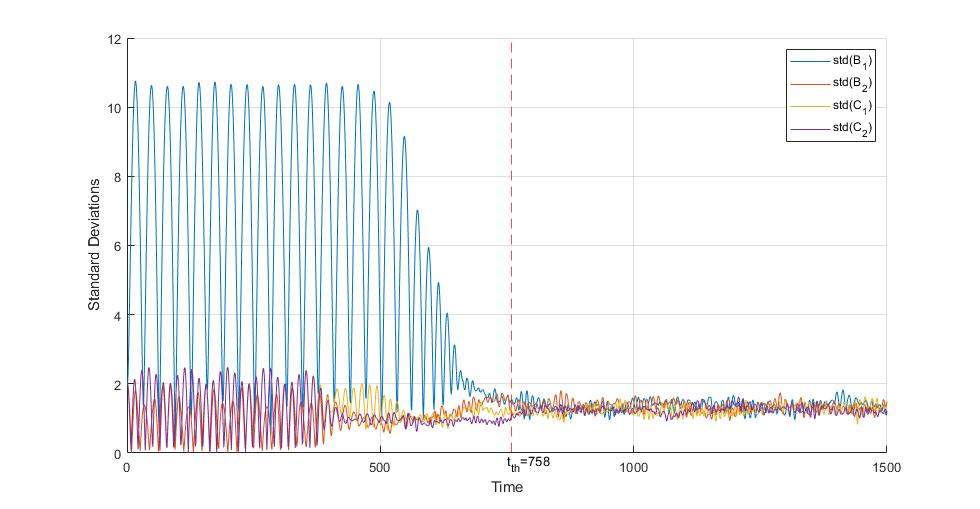}  
		\caption{Standard deviations of eigenvalues vs. Time at $N=8$, $\mu_1=1$, and $\mu_2=1.5$}
		\label{fig:fig41}
	\end{figure}

	Figure \ref{fig:fig41} displays how the  standard deviations of the eigenvalues for $B_1$, $B_2$, $C_1$, and $C_2$ matrices evolve with simulation time. 
	As seen in the legend, std($B_1$) denotes the standard deviation of the eigenvalues for $B_1$ and so on. Starting from oscillatory behavior with nearly constant      
	amplitude, std($B_1$) undergoes a change  
	at $t \gtrapprox 500$
	and its amplitude decrease considerably with time. In addition, as time progress, the standard deviations tend to converge on a narrow band of values and the system reaches a seemingly stable configuration in which only minor fluctuations are observed.  
	
	Among the different notions of thermalization time, we choose to focus on the one that define it as the timescale of thermalization from a given set of initial conditions. Using the signal processing toolbox of Matlab, we have developed a code that detects the time instants at which the variance of a signal changes significantly and run it on the standard deviation data graphed in Figure \ref{fig:fig41}. The approximate time when the standard deviations, hence the system, reaches an equilibrium size is determined to be equal to $758$. In Figure \ref{fig:fig41}, this approximate time instant is 
	marked with a dashed vertical line and 
	$t_{th}$ denotes the thermalization time of the system.
	
\begin{figure}[!htb]
		\centering
		\begin{subfigure}[!htb]{.495\textwidth}
			\centering
			\includegraphics[width= 1\linewidth]{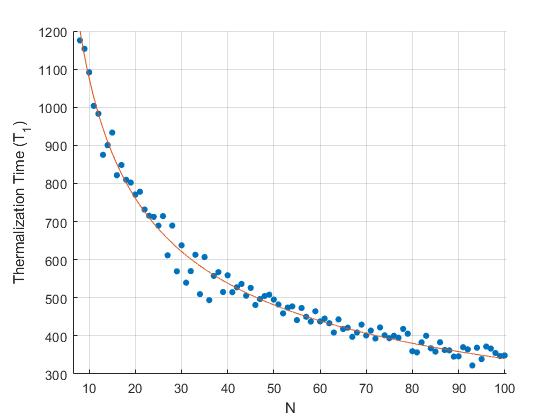}  
			\caption{$\mu_2 = 1$}
			\label{fig:fig5a1}
		\end{subfigure}	
		\begin{subfigure}[!htb]{.495\textwidth}
			\centering
			\includegraphics[width=1\linewidth]{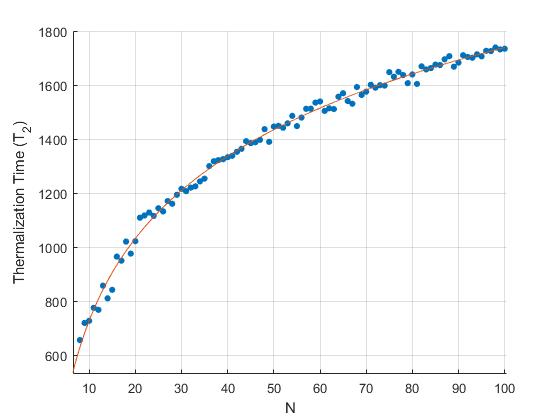}  
			\caption{$\mu_2 = 1.5$}
			\label{fig:fig5b1}
		\end{subfigure}	
		\begin{subfigure}[!htb]{.495\textwidth}
			\centering
			\includegraphics[width=1\linewidth]{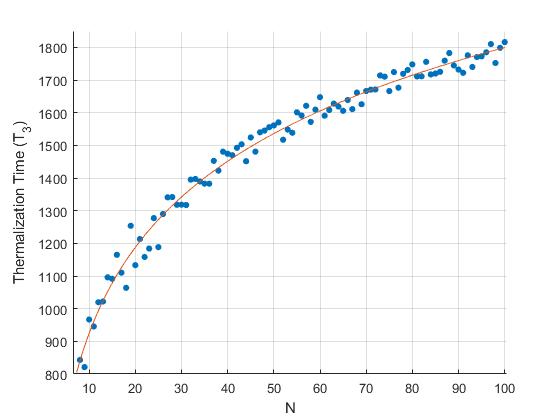}  
			\caption{$\mu_2 = 2$}
			\label{fig:fig5c1}
		\end{subfigure}	
		\begin{subfigure}[!htb]{.495\textwidth}
			\centering
			\includegraphics[width=1\linewidth]{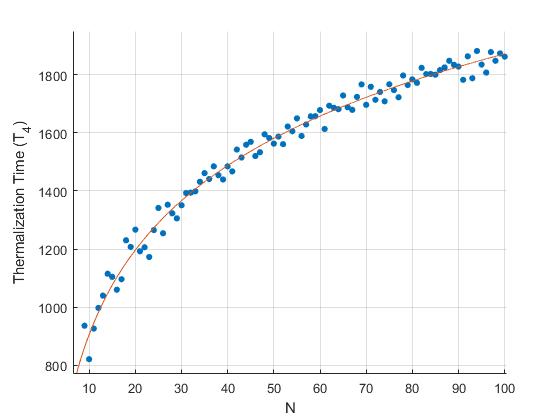}  
			\caption{$\mu_2 = 3$}
			\label{fig:fig5d1}
		\end{subfigure}	
		\caption{Thermalization time vs. $N$ at $\mu_1=1$}
		\label{fig:fig51}
	\end{figure}

	The procedure detailed above can be generalized for $N>8$. In Figure \ref{fig:fig51}, we present 
	plots of thermalization time versus $N$ at four distinct 
	mass combinations,   
	where the matrix size $N$ takes the values $N = 8,\dots,100$.
	Let us immediately note that the models at $\mu_1 = \mu_2 = 1$ have different features from the rest in the sense that data values tend to decrease with increasing $N$. We find that the function 
	\be
	\label{kokfit}
	T_1(N) = \frac{3404}{\sqrt{N}} \,,
	\ee
	provides an adequate fit 
	to the data as can be seen from Figure \ref{fig:fig5a1}. 
	In addition, 
	a logarithmic fit of the form 
	\be
	\label{logfit1}
	T_a(N) = c_a \log(N) + d_a \,,
	\ee
	with
	\begin{table}[H]
		\centering
		\begin{tabular}{ | c | c | c |}
			\cline{2-3}
			\multicolumn{1}{c |}{} &  $c_a$ & $d_a$ \\ \hline
			$T_2(N)$ &$439.5$ &$-284.9$   \\ \hline
			$T_3(N)$  &$380.1$ &$49.04$ \\ \hline
			$T_4(N)$  &$419.4$ & $-60.1$ \\ \hline
		\end{tabular}
		\caption{$c_a$ and $d_a$ values for the fitting curve (\ref{logfit1})}
		\label{table:fitvalues1}
	\end{table}
	\noindent appears to be well-suited 
	for the remaining models as can be observed from Figures \ref{fig:fig5b1} - \ref{fig:fig5d1}. In equation (\ref{logfit1}), the index $a$ ranges from $2$ to $4$. Besides, it is important to note that expressions (\ref{kokfit}) and (\ref{logfit1}) are quite sufficient to fit the data as the minimum recorded adjusted R-squared value is equal to $0.938$.  

At a slight tangent to the analysis of thermalization times, let us return back to the study of Figure \ref{fig:fig41}. The method used in the preparation of this figure can be applied with some arbitrary $p_0$ value of our choosing to produce a similar graph. In Appendix \ref{AppAddFgs}, we display Figure \ref{fig:figApp1}, which shows the variations of the standard deviations of the eigenvalues for $B_1$, $B_2$, $C_1$, and $C_2$ matrices at the $p_0$ values that are already utilized in the preparation of Figure \ref{fig:fig11}. Similar to the behavior observed when $p_0$ is equal to $30$, in Figures \ref{fig:figAppa1}-\ref{fig:figAppe1}, after periods of decrease in oscillation amplitudes, standard deviations converge on narrow bands, which implies that thermalization occurs at all six $p_0$ values. To test this hypothesis, we may pick $p_0=7$ and examine the eigenvalue distributions of the momenta matrices. In Figure \ref{fig:figAppDst}, the histograms of the eigenvalues of $P_1$ and $R_1$ at $p_0=7$ are depicted together for the sake of comparison. Since the semicircle curve provides a good fit to both distributions, we can infer that the momenta temperatures are essentially the same  and thermalization has occurred.
	
	\subsubsection{Energy dependence of thermalization time} 
	
	Apart from its dependence to matrix size, we can also explore the variation of the thermalization time with respect to energy. 
	In this subsection, by performing simulations of the matrix model (\ref{Ham1}), 
	the dependence of thermalization time to energy is depicted at several distinct mass combinations and matrix size values. 
	
	We launch the discussion with introducing a matrix configuration in the form
	\begin{align}
		\label{IniCndsFLc1} 
		B_1 = 
		\begin{pmatrix}
			v(t) J_1 & 0  \\
			0 & 0  
		\end{pmatrix} \,, \quad
		B_2 &= 
		\begin{pmatrix}
			v(t) J_2 & q_1  \\
			{q_1}^\dagger & 0  
		\end{pmatrix} \,, \quad
		B_3 = 
		\begin{pmatrix}
			v(t) J_3 & q_2  \\
			{q_2}^\dagger & 0  
		\end{pmatrix} \,,
		\nn \\
		C_l = 
		\begin{pmatrix}
			z(t) J_l & 0  \\
			0 & 0  
		\end{pmatrix} \,, \quad
		P_1 &= 
		\begin{pmatrix}
			0 & 0  \\
			0 & p_0  
		\end{pmatrix} \,, \quad
		P_2 = P_3 = 0  \,, \quad R_l = 0 \,, \nn \\
	    D_s &= 0 \,, \quad  W_s = 0 \,,
	\end{align}
	where $v(t)$ and $z(t)$ are real functions of time and $J_i$ satisfy the commutation relations given by $[J_i,J_j] = i \hbar^{}_{\!J}
	\epsilon_{i j l} J_{l}$. 
	After substituting configuration (\ref{IniCndsFLc1}), 
	at an arbitrary time $t$, into the Hamiltonian (\ref{Ham1}), we evaluate the traces using Matlab and arrive at a set of effective Hamiltonians\footnote[3]{It is important to remark that we employ this method only for producing initial configurations. Unlike the reduced models in matrix model settings studied in \cite{Baskan:2019qsb,Arefeva:1997oyf,Asano:2015eha},
	the matrix configuration defined by (\ref{IniCndsFLc1}) does not satisfy the equations of motion (\ref{HeomNw}).}.
	A generic member of this set can be expressed as follows  
	\begin{align}
	\label{EqHs}
	H_s &= \frac{1}{2} p_0^2 + {\hbar^{}_{\!J}}^{\!\!4} c^{}_{N} \Big(v^2+z^2 \Big)^2 + {\hbar^{}_{\!J}}^{\!\!2} \Big[\Big(c^{}_{N} \mu_1^2 +  \Delta_1 \Big) v^2 +
		\Big(c^{}_{N} \mu_2^2 +  \Delta_2\Big) z^2 \Big] \nn \\
	&+ \Delta_3 \mu_1^2 \,,  
	\end{align}	
 	where the coefficients $c^{}_{N}$ are defined by $c^{}_{N} = \frac{N(N-1)(N-2)}{8}$.

	Here, it is essential to note that, due to the presence of fluctuation blocks $ q_1$ and $q_2$, unlike $c^{}_{N}$, $\Delta_i$ coefficients are random numbers that change with every new substitution of the configuration (\ref{IniCndsFLc1}) into (\ref{Ham1}). With the purpose of listing and examining $\Delta_i$ values, we have repeated the procedure utilized in the obtainment of $H_s$ by running a code $500$ times and determined the reduced Hamiltonians. For $N=8$, the maximums of the absolute values of $\Delta_1$, $\Delta_2$, and $\Delta_3$ were recorded as $0.0018$, $0.0014$, $0.0003$ respectively, which indicates that the extent of change in the coefficients of quadratic terms is small (in comparison to  $c^{}_{N}$) but not negligible. Let us also add that we set $\hbar^{}_{\!J}$ to $1$ in this subsection.  
	
	Another point to emphasize is that analyzing the classical dynamics of equation (\ref{EqHs}) is not a purpose of this study. 
	$H_s$ would be solely employed to generate initial conditions for the simulations of equation (\ref{Ham1}). In order to give a detailed description of the initial condition selection process, let us first denote by $(v_b,z_b)\equiv \big(v(t_b),z(t_b)\big)$ a generic set of initial conditions at the start time $t_b$ of a classical simulation of $H$. Then, at $t=t_b$, (\ref{EqHs}) can be expressed as shown below
	\useshortskip
	\begin{align}
	E &= c^{}_{N} v_b^4 + c^{}_{N} z_b^4 + \Big(c^{}_{N} \mu_1^2 +  \Delta_1 \Big) v_b^2 + \Big(c^{}_{N} \mu_2^2 +  \Delta_2\Big) z_b^2 
	+ 2 c^{}_{N} v_b^2 z_b^2 \nn \\
	&+ \frac{1}{2} p_0^2 + \Delta_3 \mu_1^2 \,,  
	\end{align} 
   	where $E$ is the energy of the reduced action. 
	
\begin{figure}[!htb]
		\centering
		\begin{subfigure}[!htb]{.495\textwidth}
			\centering
			\includegraphics[width= 1\linewidth]{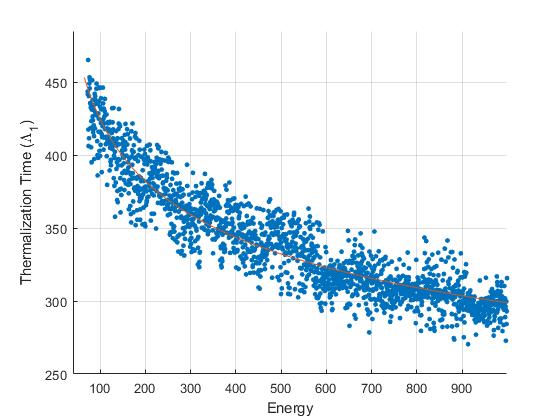}  
			\caption{$\mu_2 = 0.5$}
			\label{fig:fig6a1}
		\end{subfigure}	
		\begin{subfigure}[!htb]{.495\textwidth}
			\centering
			\includegraphics[width=1\linewidth]{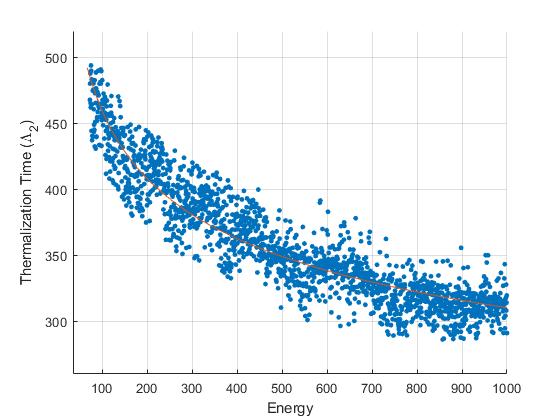}  
			\caption{$\mu_2 = 1.5$}
			\label{fig:fig6b1}
		\end{subfigure}	
		\begin{subfigure}[!htb]{.495\textwidth}
			\centering
			\includegraphics[width=1\linewidth]{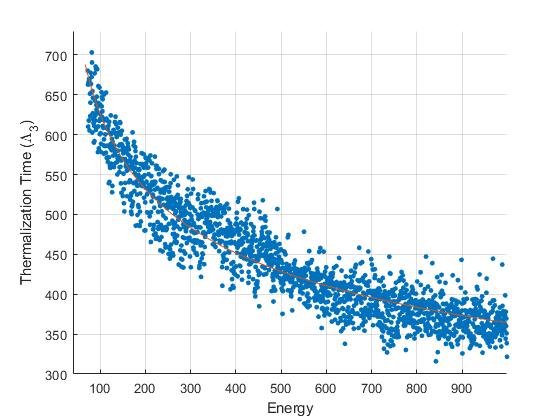}  
			\caption{$\mu_2 = 3$}
			\label{fig:fig6c1}
		\end{subfigure}	
		\begin{subfigure}[!htb]{.495\textwidth}
			\centering
			\includegraphics[width=1\linewidth]{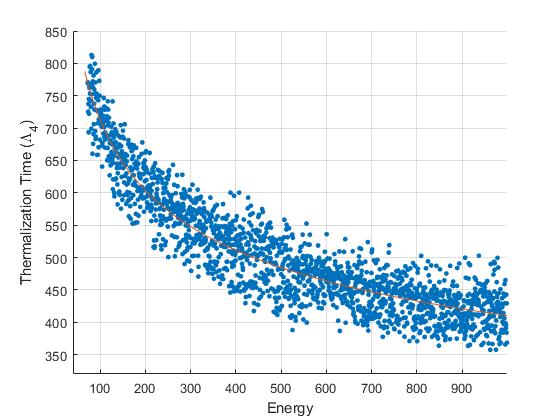}  
			\caption{$\mu_2 = 4$}
			\label{fig:fig6d1}
		\end{subfigure}	
		\caption{Thermalization time vs. Energy at $\mu_1=1$ and $N=8$}
		\label{fig:fig61}
	\end{figure}

\begin{figure}[!htb]
		\centering
		\begin{subfigure}[!htb]{.495\textwidth}
			\centering		
			\includegraphics[width=1\linewidth]{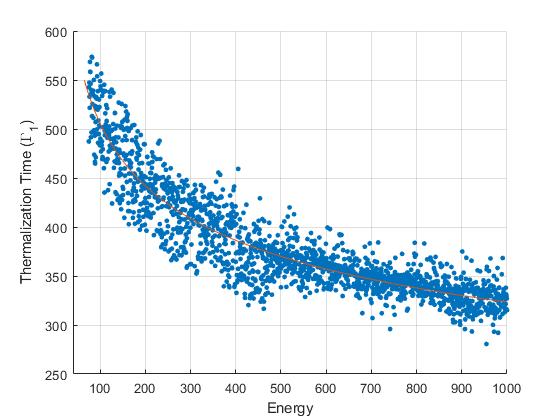}
			\caption{$N=6$}
			\label{fig:fig7a1}
		\end{subfigure}	
		\begin{subfigure}[!htb]{.495\textwidth}
			\centering
			\includegraphics[width=1\linewidth]{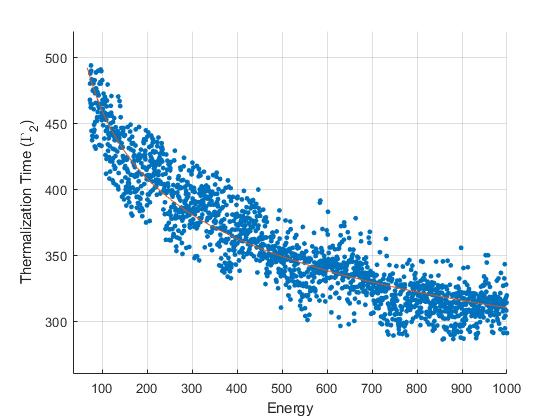}
			\caption{$N=8$}
			\label{fig:fig7b1}
		\end{subfigure}	
		\begin{subfigure}[!htb]{.495\textwidth}
			\centering
			\includegraphics[width=1\linewidth]{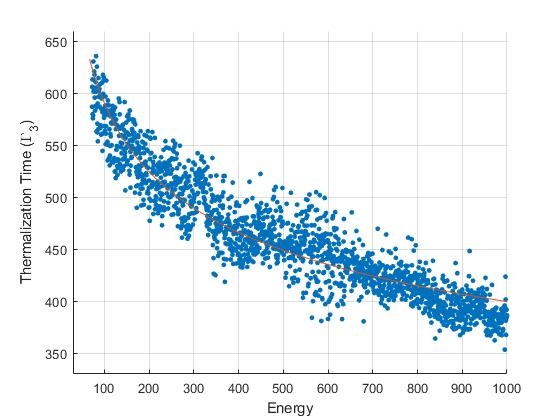} \caption{$N=10$}
			\label{fig:fig7c1}
		\end{subfigure}	
		\begin{subfigure}[!htb]{.495\textwidth}
			\centering
			\includegraphics[width=1\linewidth]{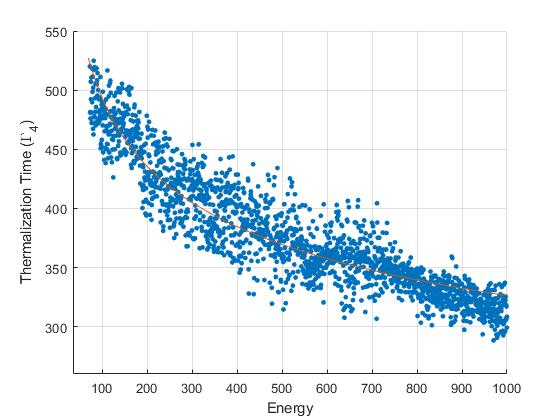} \caption{$N=12$}
			\label{fig:fig7d1}
		\end{subfigure}	
		\caption{Thermalization time vs. Energy at $\mu_1=1$ and $\mu_2=1.5$}
		\label{fig:fig71}
	\end{figure}

	With the aim of investigating the variance of thermalization time with energy, we run another Matlab code, which determines the thermalization time at several different values of the energy. We run the code with randomly selected initial conditions satisfying a given energy condition and detect the thermalization time of the system for a specified matrix size and mass combination. In order to give certain effectiveness to the random initial condition selection process, we developed a simple approach which we briefly explain next. 	To start with, we generate two uniformly distributed random numbers $\phi_\nu$ over the interval $O \leq \phi_\nu \leq E$ satisfying the constraint   
	$E = \phi_1 + \phi_2$ . Subsequently, the real roots of the expression 
	\useshortskip
	\be
	\label{iniCon1}
	c^{}_{N} z_b^4 + \big( c^{}_{N} \mu_2^2 + \Delta_2 \big) z_b^2 
	+ \frac{1}{2} {p_0}^2 + \Delta_3 \mu_1^2 - \phi_1 = 0 \,,
	\ee
	are found. Our code randomly selects one of these roots, which is later used to solve for $v_b$ in the equation 
	\useshortskip
	\be
	\label{iniCon2}
	c^{}_{N} v_b^4 + \big( c^{}_{N} \mu_1^2 +2 c^{}_{N} z_b^2
	+ \Delta_1 \big) v_b^2 - \phi_2 = 0 \,.
	\ee
	Lastly, as the final step of the selection process, one of the real roots of equation (\ref{iniCon2}) is randomly picked by our code. Having now determined the $(v_b,z_b)$ pair, we move on to discuss the simulation stage. In order to measure the thermalization time at the energy $E$, we perform a classical simulation of the matrix model (\ref{Ham1}). This simulation is started with the initial configuration given by
		\begin{align}
		B_1 = 
		\begin{pmatrix}
			v_b J_1 & 0  \\
			0 & 0  
		\end{pmatrix} \,, \quad
		B_2 &= 
		\begin{pmatrix}
			v_b J_2 & q_1  \\
			{q_1}^\dagger & 0  
		\end{pmatrix} \,, \quad
		B_3 = 
		\begin{pmatrix}
			v_b J_3 & q_2  \\
			{q_2}^\dagger & 0  
		\end{pmatrix} \,,
		\nn \\
		C_l = 
		\begin{pmatrix}
			z_b J_l & 0  \\
			0 & 0  
		\end{pmatrix} \,, \quad
		P_1 &= 
		\begin{pmatrix}
			0 & 0  \\
			0 & p_0  
		\end{pmatrix} \,, \quad
		P_2 = P_3 = 0  \,, \quad R_l = 0 \,, \nn \\
		D_s &= 0 \,, \quad  W_s = 0 \,.
	\end{align}
	Following the completion of the classical simulation,
	the thermalization time is measured by the method described at the beginning of subsection \ref{Ssec_Thrm}.
	By setting $p_0$ equal to $12$ and repeating the procedure detailed above for a range of energy values, 
	the data used in the depiction of
	Figures \ref{fig:fig61} and \ref{fig:fig71} are prepared. 
 
	Figure \ref{fig:fig61} shows the plots of thermalization time versus energy at four different $\mu_2$ values.  
	The best-fitting function for the numerical data displayed in Figure \ref{fig:fig61} is found to be  
	a power law in the form  
	\be
	\label{pwrFit1}
	\Lambda_m(E) = \alpha_m E^{\beta_{m}} + \xi_m \,.
	\ee
	The fitting parameters of the best fit equations (\ref{pwrFit1}) are listed in Table \ref{tab:pwFitvals1}. Due to the obvious increase in the variance of the data, the fits describing the thermalization times of 
	\begin{table}[H]
		\centering
		\begin{tabular}{ | c | c | c | c |}
			\cline{2-4}
			\multicolumn{1}{c |}{} &  $\alpha_m$ & $\beta_m$ & $\xi_m$ \\ \hline
			$\Lambda_1(E)$  &$859.8$ &$-0.1491$ & $-8.2$   \\ \hline
			$\Lambda_2(E)$  &$1007$ &$-0.1737$ & $6.9$ \\ \hline
			$\Lambda_3(E)$  &$1850$ & $-0.2353$ & $0.4$ \\ \hline
			$\Lambda_4(E)$  &$2153$ & $-0.240$ & $3.1$ \\ \hline
		\end{tabular}
		\caption{$\alpha_m$, $\beta_m$ and $\xi_m$ values for the fitting curve (\ref{pwrFit1}) \label{tab:pwFitvals1}}
	\end{table}
	\noindent Figure \ref{fig:fig61} are not as good in comparison to the fits displayed in Figure \ref{fig:fig51}. The four fitting curves $\Lambda_m$ ($m = 1,2,3,4$) appear to have the adjusted R-squared statistics of $0.8681$, $0.8654$, $0.897$, and $0.8703$ respectively.
	
	On the other hand, in order to take the effects of matrix size into consideration, we illustrate in Figure \ref{fig:fig71} the evolutions of thermalization times with energy at $N=6,8,10,12$. From the profile of thermalization times with respect to energy shown in Figure \ref{fig:fig71}, we observe that numerical data exhibits a decreasing trend, which can be modelled again with a power law in the form 
	\be
	\label{pwrFit2}
	\Gamma_m(E) = \theta_m E^{\epsilon_{m}} + \delta_m \,,
	\ee
	with the fitting parameters displayed in Table \ref{tab:pwFitvals2}. The adjusted R-squared values of the 
	\begin{table}[H]
		\centering
		\begin{tabular}{ | c | c | c | c |}
			\cline{2-4}
			\multicolumn{1}{c |}{} &  $\theta_m$ & $\epsilon_{m}$ & $\delta_m$ \\ \hline
			$\Gamma_1(E)$  &$1237$ &$-0.1942$ & $-3.4$ \\ \hline
			$\Gamma_2 (E)$ &$1007$ &$-0.1737$ & $6.9$   \\ \hline
			$\Gamma_3(E)$  &$1291$ & $-0.1825$ & $33.7$ \\ \hline
			$\Gamma_4(E)$  &$1128$ & $-0.1796$ & $1.8$ \\ \hline
		\end{tabular}
		\caption{$\theta_m$, $\epsilon_m$ and $\delta_m$ values for the fitting curve (\ref{pwrFit2}) \label{tab:pwFitvals2}}	
	\end{table}
	\noindent fitting curves depicted in Figures \ref{fig:fig7a1} - \ref{fig:fig7d1} are given by $0.8548$, $0.8654$, $0.8722$ and $0.855$ respectively, which essentially indicates that $\Gamma_m$ curves provide adequate fits to the numerical data.
	
	\section{Conclusions and outlook} \label{Concs}
	
	In this paper, we have considered the dynamics of thermalization in a Yang-Mills matrix model with two distinct mass deformation terms, which may be contemplated as a double mass deformation of the bosonic part of the BFSS model. We have performed a detailed numerical analysis 
	of the 
	classical evolution of this model and determined that when the simulations are started from a certain set of initial conditions, thermalization occurs. Although small background fluctuations are required to initiate thermalization, from the findings 
	of numerical simulations it was clearly seen that thermalization times are independent of these fluctuations. 
	This is an extension of the result given in \cite{Riggins:2012qt} for the BFSS model. 
	
	From the results concerning the change in thermalization times with respect to matrix size,
	we were able to demonstrate through an appropriate fitting function that thermalization times vary logarithmically with 
	matrix size when the mass parameters $\mu_1$ and $\mu_2$ differ. 
	It is worth mentioning that in reference \cite{Sekino:2008he}, thermalization (or scrambling) time of a black hole is conjectured to be proportional to $\log(N)$ where $N$ is the number of degrees of freedom. Even though 
	we have adopted a different definition of thermalization time, it is still interesting to note that 
	the findings obtained for Hamiltonian (\ref{Ham1}) 
	confirm this conjecture. In subsection \ref{Ssec_Thrm}, we have also presented plots depicting the variations of thermalization times with 
	respect to 
	the energies of the reduced actions and subsequently the best-fitting functions for the data were determined as power laws. A common feature observed in all fitting functions is that 
	thermalization times converge to finite values in the large energy or matrix size limit.  
	
	Let us also mention some recent developments in related subjects.
	Although calculating entanglement entropy in ordinary field theories is a rather difficult task, calculations in noncommutative theories such as the scalar field theory on the fuzzy sphere were already carried out in \cite{Dou:2006ni, Karczmarek:2013jca, Okuno:2015kuc}. Moreover, numerical computations of entanglement entropy in the BFSS matrix model were recently performed in \cite{Buividovich:2018scl}.
	Besides, the behavior of entanglement entropy during thermalization was
	studied in holographic systems in references \cite{Liu:2013qca, Arefeva:2017pho, Arefeva:2020uec}. Based on these considerations, a valuable direction of research would be to investigate the time dependence of entanglement entropy in the system defined by (\ref{Ham1}). Particularly, it would be interesting to explore the possible use of entanglement entropy as a probe of thermalization.
	Another challenging direction of development is to analyze the dynamics of quantum chaos with emphasis on the measurements of Lyapunov exponents and check whether our model saturates the Maldacena-Shenker-Stanford bound \cite{Maldacena:2015waa} or not. We hope that these issues will produce useful results to be reported soon.

\appendix

\section{Additional figures}\label{AppAddFgs}

\begin{figure}[!htb]
	\centering
	\includegraphics[width= 0.62\linewidth]{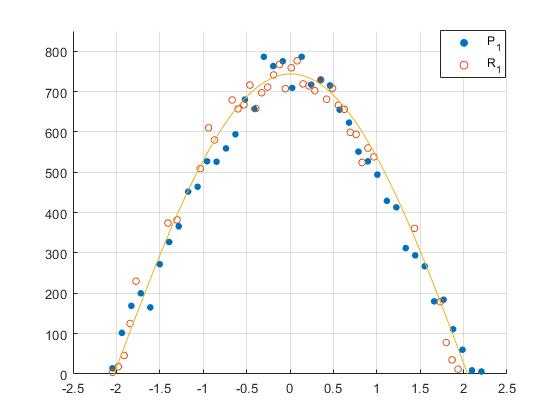}  
	\caption{Histograms of eigenvalues of $P_1$ and $R_1$ at $N=8$, $p_0 = 7$, $\mu_1=1$, and $\mu_2=1.5$}
	\label{fig:figAppDst}
\end{figure}

In this appendix, we present Figures \ref{fig:figAppDst} and \ref{fig:figApp1}. We illustrate in
Figure \ref{fig:figAppDst} the eigenvalue distributions of  $P_1$ and $R_1$ at $p_0 = 7$. The histograms are generated by sampling the momenta eigenvalues on the time interval $[827,3000]$ with a bin size equal to $40$. In Figure \ref{fig:figApp1}, the time evolutions of the standard deviations of the eigenvalues for $B_1$, $B_2$, $C_1$, and $C_2$ matrices 
at six different $p_0$ values are displayed. 
\begin{figure}[!htb]
	\centering
	\begin{subfigure}[!htb]{.495\textwidth}
		\centering
		\includegraphics[width= 1\linewidth]{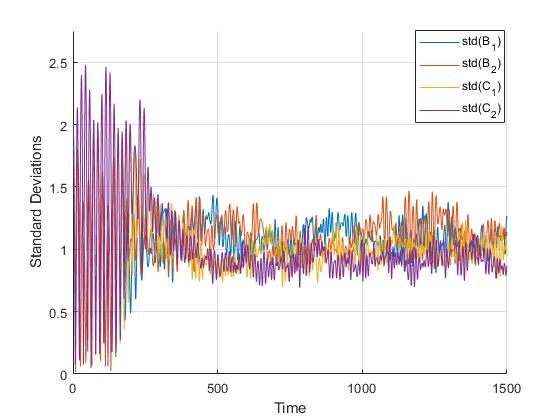}  
		\caption{$p_0 = 0$}
		\label{fig:figAppa1}
	\end{subfigure}	
	\begin{subfigure}[!htb]{.495\textwidth}
		\centering
		\includegraphics[width=1\linewidth]{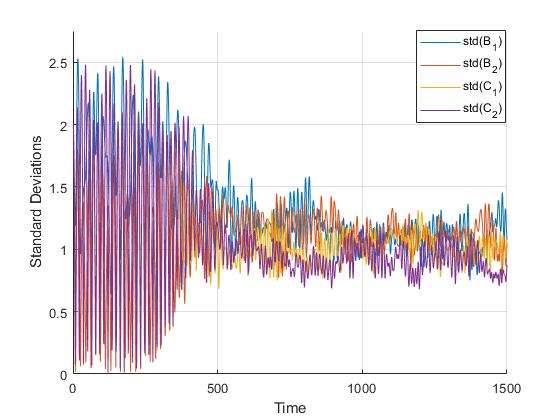}  
		\caption{$p_0 = 5$}
		\label{fig:figAppb1}
	\end{subfigure}	
\end{figure}
\begin{figure}[!htb]\ContinuedFloat
	\begin{subfigure}[!htb]{.495\textwidth}
		\centering
		\includegraphics[width=1\linewidth]{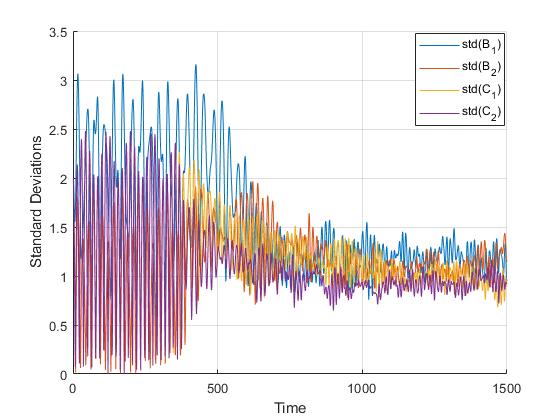}  
		\caption{$p_0 = 7$}
		\label{fig:figAppc1}
	\end{subfigure}	
	\begin{subfigure}[!htb]{.495\textwidth}
		\centering
		\includegraphics[width=1\linewidth]{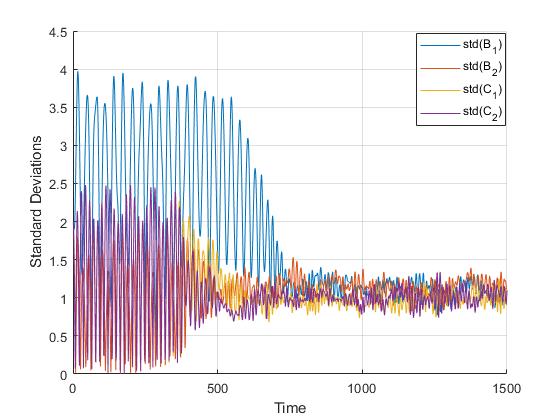}  
		\caption{$p_0 = 10$}
		\label{fig:figAppd1}
	\end{subfigure}	
	\begin{subfigure}[!htb]{.495\textwidth}
		\centering
		\includegraphics[width= 1\linewidth]{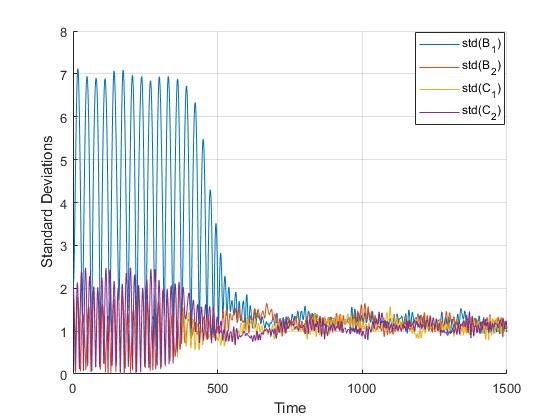}  
		\caption{$p_0 = 19.5$}
		\label{fig:figAppe1}
	\end{subfigure}	
	\begin{subfigure}[!htb]{.495\textwidth}
		\centering
		\includegraphics[width=1\linewidth]{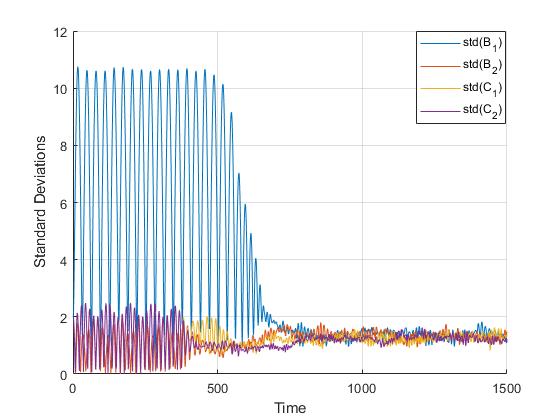}  
		\caption{$p_0 = 30$}
		\label{fig:figAppf1}
	\end{subfigure}	
	\caption{Standard deviations of eigenvalues vs. Time at $N=8$, $\mu_1=1$, and $\mu_2=1.5$}
	\label{fig:figApp1}
\end{figure}

\end{document}